\documentclass[preprint,superscriptaddress,showpacs,preprintnumbers,amsmath,amssymb,aps,prb,floatfix,]{revtex4-1}

\usepackage{graphicx}
\usepackage{dcolumn}
\usepackage{bm}
\usepackage{hyperref}
\usepackage{color,soul} 

\begin{document}
\title{Giant Enhancement of Intrinsic Spin Hall Conductivity in $\beta$ Tungsten via Substitutional Doping} 

\author{Xuelei Sui}
\affiliation{Department of Physics and Astronomy, California State University Northridge, Northridge, California 91330, USA}
\affiliation{Department of Physics and State Key Laboratory of Low-Dimensional Quantum Physics, Tsinghua University, Beijing 100084, China}
\affiliation{Computational Science Research Center, Beijing 100084, China}

\author{Chong Wang}
\affiliation{Institute for Advanced Study, Tsinghua University, Beijing 100084, China}

\author{Jinwoong Kim}
\affiliation{Department of Physics and Astronomy, California State University Northridge, Northridge, California 91330, USA}

\author{Jianfeng Wang}
\affiliation{Department of Physics and State Key Laboratory of Low-Dimensional Quantum Physics, Tsinghua University, Beijing 100084, China}
\affiliation{Computational Science Research Center, Beijing 100084, China}

\author{S. H. Rhim}
\affiliation{Department of Physics and Energy Harvest Storage Research Center, University of Ulsan, Ulsan 680-749, Korea}

\author{Wenhui Duan}
\email{dwh@phys.tsinghua.edu.cn}
\affiliation{Department of Physics and State Key Laboratory of Low-Dimensional Quantum Physics, Tsinghua University, Beijing 100084, China}
\affiliation{Institute for Advanced Study, Tsinghua University, Beijing 100084, China}

\author{Nicholas Kioussis}
\email{nick.kioussis@csun.edu}
\affiliation{Department of Physics and Astronomy, California State University Northridge, Northridge, California 91330, USA}

\date{\today}

\begin{abstract}
A key challenge in manipulating the magnetization in heavy-metal/ferromagnetic bilayers via the spin-orbit torque is
to identify materials that exhibit an efficient charge-to-spin current conversion. {\it Ab initio} electronic structure
calculations reveal that the intrinsic spin Hall conductivity (SHC)
for pristine $\beta$-W is about sixty percent larger than that of
$\alpha$-W. More importantly, we demonstrate that the SHC of $\beta$-W can be enhanced via Ta alloying. This is corroborated by spin Berry curvature calculations of W$_{1-x}$Ta$_x$ ($x$ $\sim$ 12.5\%) alloys which show a giant enhancement of spin Hall angle of up to $\approx$ $-0.5$. The underlying mechanism is the synergistic behavior of the SHC and longitudinal conductivity with Fermi level position. These findings, not only pave the way for enhancing the intrinsic spin Hall effect in $\beta$-W, but also provide new guidelines to exploit substitutional alloying to tailor the spin Hall effect in various materials.

\end{abstract}


\maketitle

The current-driven spin-orbit torques (SOT) in heavy-metal/ferromagnetic heterostructures
have drawn increasing attention because they can
provide an efficient way of manipulating the magnetization\cite{Miron2010,Liu2012,Liu2011}.
The SOT arises from the transfer of spin angular momentum between charge current and the local magnetization in the presence of spin-orbit coupling (SOC). The underlying origin of the SOT is still under debate: it may arise from either
the bulk spin Hall effect (SHE)\cite{Dyakonov1971,Hirsch1999,Jungwirth2012,Sinova2015} in the heavy metal
or the interfacial SOC\cite{Kim2012,Kurebayashi2014} or both\cite{Freimouth2014,Fan2014}. In the SHE, an
in-plane charge current density, $J_c$, flowing in the heavy metal with strong SOC generates
a transverse spin current, $J_s$.
The SHE efficiency and the critical current density for magnetization switching is expressed in terms of the spin Hall angle (SHA), $\theta_{SH} = J_s/J_c$\cite{Dyakonov1971,Hirsch1999}, which measures the efficiency
of the charge current to spin current conversion. It only depends on the material properties and can be experimentally measured  via the inverse spin Hall effect\cite{ISHE,ISHE2}, the spin Seebeck effect\cite{Seebeck,Seebeck2}, the Kerr effect\cite{Kerr}, or the spin Nernst effect\cite{Nernst}.

Relatively large values of $\theta_{SH}$ have been reported in the 5$d$ elemental solids of Pt (0.07)\cite{Liu2012,Azevedo2011} and the
high-resistive $\beta$-Ta ($-0.15$)\cite{Liu2012}. More recently, very thin tungsten films in the highly resistive ($\rho_{\beta-W} \sim$ 100--300 $\mu\Omega\cdot$cm) metastable
$\beta$-phase (A15 crystal structure) were reported to exhibit giant spin Hall effect with a $\theta_{SH}$ $\sim -0.3$ to $-0.4$\cite{Pai2012,Hao2015a,Hao2015b}, the
largest spin Hall angle among simple element transition metals. In sharp contrast, thicker and/or annealed W films form in the relatively low-resistive ($\rho_{\alpha-W} \leq$ 25 $\mu\Omega\cdot$cm) $\alpha$-W phase ($bcc$ crystal structure) exhibit small ($>$ $-0.07$) spin Hall angle\cite{Pai2012}. These results invite the intriguing question what is the origin in the electronic structure of the dramatically different intrinsic spin Hall conductivities
and the spin Hall angles of the $\beta$- and $\alpha$-W.

In this rapid communication, we compare the intrinsic spin Hall conductivity (SHC) of bulk W
in the $\alpha$- (bcc) and $\beta$- (A15) phases using first-principles calculations.
The calculations reveal that both phases possess large SHC,
where the {\em k}-resolved spin Berry curvature elucidates the resonant feature of the spin orbit splitting of the doubly degenerate bands near the Fermi level.
For $\beta$-W, the large SHC along with the large resistivity
work cooperatively to yield large SHA.
We propose the `acceptor' alloying in $\beta$-W can further enhance the SHA from the analysis of electronics structure.
More specifically, Ta substitution locates the Fermi level locate inside the SOC induced bandgap, rendering the SHC remarkably gigantic, which suggests a way to increase the SHE.

{\it Ab initio} electronic structure calculations have been carried out within the framework of the projector augmented-wave formalism\cite{Blochl94},
as implemented in the Vienna ab initio simulation package (VASP)\cite{Kresse96a,Kresse96b,KressePAW}.
The generalized gradient approximation is used to describe the exchange-correlation functional as parameterized by Perdew-Burke-Ernzerhof\citep{PBE}.
The plane-wave cutoff energy is 500 eV and the first Brillouin zone (BZ) is sampled using a $\Gamma$-centered special k-point grid with $24 \times 24 \times 24$ for $\alpha$-W and $14 \times 14 \times 14$ for $\beta$-W, respectively. All structures are fully relaxed with force criteria of 0.001 eV/{\AA}.
The SOC of the valence electrons is in turn included using the second-variation method\cite{Koelling} employing the scalar-relativistic eigenfunctions
of the valence states.
Then DFT wave functions were projected to maximally-localized Wannier functions using the Wannier90 package\cite{Wannier,Wannier90,Marzari2001} and
 the Kubo formula\cite{kubo} is employed to calculate the SHC.
A dense {\em k}-mesh of $100 \times 100 \times 100$ and $90 \times 90 \times 90$ on the full BZ for $\alpha$-W and $\beta$-W, respectively, is employed
 to perform the BZ integration for the intrinsic SHC
 because of the slow convergence caused by the large contributions of both signs to the spin Berry curvature
which occur in very small regions of $k$ space.
The longitudinal charge conductivity is calculated using the Boltzmann transport equation\cite{boltzman2014}.

In the clean case (impurity potential $V$=0), the intrinsic SHC within the Kubo-formalism involves an integration of the spin Berry curvature, $\Omega_n^z(\bm {k})$, of the occupied bands over the BZ\cite{Sinova2015,Niu2005}
\begin{equation}
  \sigma_{ab}^{c} =e{\hbar}\int_{\text{BZ}}{\frac{d\bm{k}}{(2\pi)^3}\sum\limits_n{{{f}_{{\bm{k}}n}}\Omega_{n,ab}^{c}(\bm{k})}},
  \label{eq:sigma_z}
\end{equation}
where $f_{\bm{k}n}$ is the Fermi-Dirac distribution function for the $n$-th band at $\bm{k}$ and
the Berry curvature of the $n$-th band is
\begin{equation}
\Omega _{n,ab}^{c}(\bm{k})  =-\sum\limits_{n'\ne n}\frac{2{\text{Im}}[\langle {\bm{k}}n|\hat j_{a}^{c}|{\bm k}n'\rangle \langle{\bm{k}}n'|{\hat {v}_{b}}|{\bm{k}}n\rangle]}{({\varepsilon }_{\bm{k}n}-{\varepsilon }_{\bm{k}n'})^2}.
  \label{SHC}
\end{equation}
Here, $j_{a}^{c}=\frac{1}{2} \left \{ s^c,v_a \right \}$, is the spin current operator,
$a$, $b$, and $c$ denote the three Cartesian directions $x$, $y$ and $z$,
${{s}^{c}}=(\hbar /2)\beta {{\Sigma^c}}$, where, $\beta$ is $4\times 4$ matrix and ${\Sigma^c}$ is the spin
operator in the Dirac equation, and  $|{\bm k}n\rangle$ represents the periodic part of the Bloch wave
function with energy $\varepsilon_{\bm{k}n}$. Note that the spin Berry curvature can be large near the band degeneracy.
The second-rank SHC tensor, $\sigma_{ab}^c$, describes the spin current ($J_s$) generated along the $a$th direction with spin polarization along $c$ due to
a charge current ($J_e$) flowing along the $b$th direction.
Both $\alpha$- and $\beta$-W crystallize in the centrosymmetric cubic lattices:
the $\alpha$-phase in the body-centered-cubic ({\em bcc}) with space group \emph{Im$\bar{3}$m} (\#229) while the
$\beta$-phase in the so-called A15 structure with space group \emph{Pm$\bar{3}$n} (\#223).
Consequently, $\sigma_{xy}^z = -\sigma_{yx}^z = \sigma_{yz}^x = -\sigma_{zy}^x = \sigma_{zx}^y = -\sigma_{xz}^y$ by permutation symmetry.

There has been a longstanding issue of the SHC cancellation in the clean limit
(impurity potential $V \rightarrow$ 0) due to the vertex corrections\cite{Inoue2004}. We expect that the SHC in the clean limit ($V \rightarrow$ 0) is given by the intrinsic SHC value ($V$ = 0) due to the vanishing vertex corrections under the symmetry of $H({\bm k})$ = $H({-\bm k})$\cite{Murakami2004,Murakami2008}.

\begin{figure} [tbp]
\centering
\includegraphics[width=0.7\textwidth]{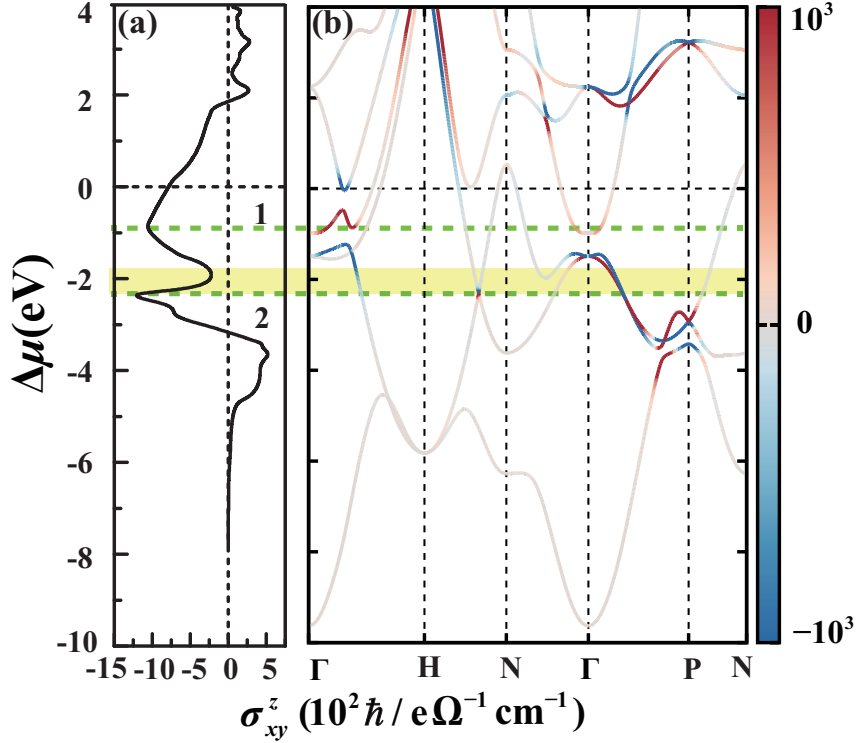}\\
\caption{(Color online) (a) Spin Hall conductivity as a function of Fermi level position, $\Delta \mu$; and (b) relativistic band structure along symmetry directions of $\alpha$-W. The color denotes the spin Berry curvature, $\Omega _{n}^{z}(\bm{k})$, of the $n$th band for each k-point which ranges from negative (blue) to positive (red). The color bar is in arbitrary units. The two green lines through (a) and (b) label the two peaks of SHC. The yellow shaded region denotes the range of $\Delta \mu$ where the SHC displays the largest decrease. The zero energy and the dashed horizontal line is the Fermi level.}
 \label{fig:Figure1}
\end{figure}

\textit{$\alpha$-W  ---} Figure~\ref{fig:Figure1} shows the relativistic band structure and the SHC ($\sigma_{xy}^z$) as function of Fermi level position, $\Delta \mu = \mu -E_F$, ($E_F$ is the Fermi level) for $\alpha$-W.
At $\Delta\mu = 0$,  $\sigma_{xy}^z = -785$ ($\hbar/e$) ($\Omega\cdot$cm)$^{-1}$. Its absolute value is smaller than that in Pt [2,200 ($\hbar/e$) ($\Omega\cdot$cm)$^{-1}$]\cite{Murakami2008} but larger than that in Au [400 ($\hbar/e$) ($\Omega\cdot$cm)$^{-1}$]\cite{Guo2009}. The magnitude of SHC decreases monotonically as $\Delta\mu$, is raised to about +2.0 eV where the SHC reverses sign. On the other hand, when $\Delta\mu$ is lowered the magnitude of the SHC increases considerably,
reaching the value of $-1,058$ ($\hbar/e$)($\Omega\cdot$cm)$^{-1}$ at $E_F = -0.8$ eV (indicated by the
green dotted line labeled by 1. The absolute value of $\sigma_{xy}^z$ reaches its maximum value of 1,206 ($\hbar/e$) ($\Omega\cdot$cm)$^{-1}$ at $\Delta\mu = -2.4$ eV as denoted by the green line (label 2) in Fig.~\ref{fig:Figure1}.
Further lowering of $\Delta\mu$ results in a decrease of SHC which eventually becomes small for $\Delta\mu< -5.0$ eV, a consequence associated with the vanishing SOC of the 5$s$ orbital.

In order to elucidate the origin of the SHC change, we show in Fig.~\ref{fig:Figure1}(b) the
{\em k}-resolved spin Berry curvature, $\Omega _{n}^{z}(\bm{k})$, of the $n$th band, where red (blue) denotes positive (negative) contribution. As shown clearly, double degeneracies which are lifted by spin-orbit coupling result in large SHC. Below $-3.0$ eV, the dominant contribution arises from regions in the vicinity of the high symmetry point $P$ with $\Omega _{n}^{z}(\bm{k})>$ 0. As the energy increases, the largest contributions to $\Omega _{n}^{z}(\bm{k})$ arise from the $\Gamma$-$P$ symmetry direction with coexistence of positive and negative values, resulting in $\sigma_{xy}^z = -1,206$ ($\hbar/e$) ($\Omega\cdot$cm)$^{-1}$ for $\Delta \mu = -2.4$ eV. As the Fermi level position further rises,
the sign of $\Omega _{n}^{z}(\bm{k})$ along $\Gamma$-H direction reverses, leading to another SHC peak
at about $-0.8$ eV and then its magnitude decreases again.


\begin{figure} [tbp]
\centering
 \includegraphics[width=0.7\textwidth]{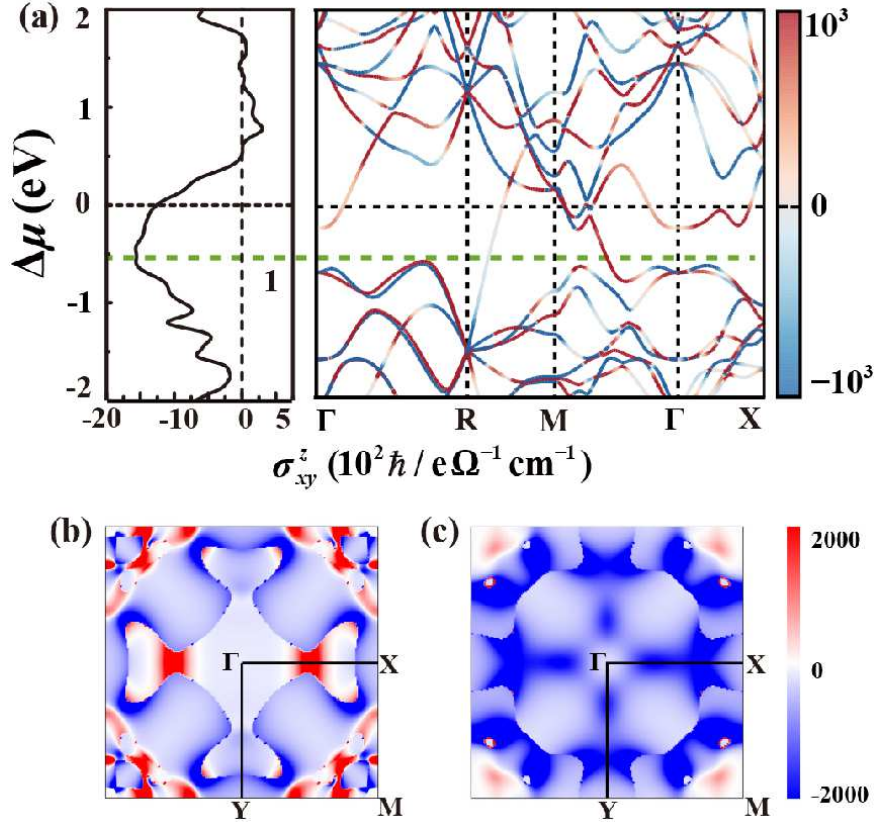}\\
 \caption{(Color online) (a) Spin Hall conductivity as a function of Fermi level position, and relativistic band structure along symmetry directions of $\beta$-W. The color denotes the spin Berry curvature, $\Omega _{n}^{z}(\bm{k})$, of the $n$th band for each k-point which ranges from negative (blue) to positive (red). The color bar is in arbitrary units. The horizontal green line denotes $\Delta\mu$ at which the SHC is maximum. The $\bf{k}_\|$-resolved spin Berry curvature in the 2D BZ ($k_x,k_y$) at $k_z=0$ for the Fermi level position at (c) 0 eV and (d) $-0.4$ eV, respectively.}
 \label{fig:Figure2}
\end{figure}

\textit{$\beta$-W  ---} Figure~\ref{fig:Figure2}(a) shows
the band structure and the SHC ($\sigma_{xy}^z$) as a function of $\Delta\mu$ for $\beta$-W.
Interestingly, the SHC is $-1,255$ ($\hbar/e$) ($\Omega\cdot$cm)$^{-1}$ at $\Delta\mu$ = 0, which is 60\% larger than that of $\alpha$-W. The variation of SHC with the Fermi level position is more
complicated compared to that for the $\alpha$ case due to the more complex band structure.
Nonetheless, $\sigma_{xy}^z$  increases as $\Delta\mu$ is lowered, reaching its maximum value
of $-1,565$ ($\hbar/e$) ($\Omega\cdot$cm)$^{-1}$ at $E_F = -0.4$ eV [green horizontal line 1 in Fig.~\ref{fig:Figure2}(a)]
due to the elimination of the large positive contribution of Berry curvature in the energy window from $-0.4$ to 0 eV. Remarkably, when the SHC reaches its maximum value, the Fermi level lies in the gap along $\Gamma-X$, where the band degeneracy is lifted by spin-orbit coupling. Figures ~\ref{fig:Figure2}(b) and (c) show the $\bf{k}_\|$-resolved spin Berry curvature in the 2D BZ ($k_x,k_y$) at $k_z = 0$ for the Fermi level position at 0 eV and $-0.4$ eV, respectively. As expected, the Berry curvature depends sensitively on $\Delta\mu$  including a sign reversal throughout large fraction of the BZ, especially along the $\Gamma$-$X$ and around the $Y$-$M$ directions. These changes and the emergence of overwhelming negative spin Berry curvature in Fig.~\ref{fig:Figure2}(c) are consistent with the band structure of $\beta$-W.

\textit{Temperature dependence of SHC ---}
The temperature-dependence of the SHC enters through the Fermi-Dirac function, $f_{{\bf k},n}$, in Eq.~(\ref{eq:sigma_z}).
We find weak temperature-dependence of the SHC in both the $\alpha$ and $\beta$ phases (change less than 2\% from low to high temperature) [see Fig. S1 in Supplementary Information\cite{support}], in sharp contrast to the strong temperature dependence reported
in Pt\cite{Murakami2008}, Pd\cite{Guo2009} and the TaAs Weyl semimetal\cite{Felser2016}.
The high value of SHC in tungsten which can be retained even at high temperature offers a great advantage for room-temperature applications.

\textit{Spin Hall Angle ---}
The SHA can be expressed as
\begin{flalign}
   {\theta _{SH}}  = \frac{e}{\hbar }\frac{{\sigma _{xy}^z}}{{{\sigma _{xx}}}}
  \label{SHA}
\end{flalign}
where $\sigma_{xx}$ is the longitudinal charge conductivity, $\sigma_{xy}^{z}$ is the transverse spin conductivity (SHC),
and the prefactor $\frac{e}{\hbar}$ renders $\theta_{SH}$ dimensionless.
Obviously, to obtain large SHA, large SHC as well as low charge conductivity are equally important.

In order to determine the SHA we have also calculated the longitudinal conductivity, $\sigma_{xx}$,
using the Boltzmann transport equations within the constant relaxation time approximation\cite{boltzman2014}.
According to the experimental resistivity values of 25 $\mu\Omega\cdot$cm and 300 $\mu\Omega\cdot$cm for $\alpha$- and $\beta$-W
\cite{Hao2015a,resistivity,resistivity2,resistivity3}, we find that the corresponding room-temperature relaxation times
are 5.52 $fs$ and 1.61 $fs$, respectively. Assuming that the relaxation time is independent of
$\Delta\mu$, we show in Figs. S2(a)(b) (see Supplementary Information\cite{support}) the variation of the calculated $\sigma_{xx}$ as a function
of $\Delta \mu$ for $\alpha$- and $\beta$-W, respectively.
In addition to the larger spin Hall current in $\beta$-W, its high resistivity leads to lower Joule heating for generation of the same amount of spin Hall current.
The calculations also reveal that lowering $\Delta \mu$ $\sim -0.2$ eV leads to a much larger increase of $\sigma_{xx}$ compared
to the corresponding increase of SHC [Fig.~\ref{fig:Figure1}(a)] in $\alpha$-W which in turn reduces the SHA to about $-0.01$. In sharp contrast, lowering $\Delta \mu$  $\sim -0.2$ eV in $\beta$-W decreases $\sigma_{xx}$ while increases the SHC, resulting in a 47\% enhancement of the SHA value ($\sim -0.47$) over its value of $-0.32$ at $\Delta\mu = 0$ eV, suggesting that the SHA can be enhanced by Fermi level lowering (see Figs. S3(a)(b) in Supplementary Information\cite{support}). Table I summarizes values of the SHC at $\Delta\mu = 0$ eV, the maximum SHC$_{\max}$, the SHA at $\Delta\mu$ = 0 eV, and the maximum SHA$_{\max}$ for the $\alpha$- and $\beta$-W structures, respectively.
	
\begin{table}[tbp]
	\caption{Calculated values of the SHC at $\Delta\mu = 0$ eV, the maximum SHC$_{\max}$, the SHA at $\Delta\mu$ = 0 eV, and the maximum SHA$_{\max}$ for $\alpha$- and $\beta$-W structures, respectively.
	Experimental values of the SHA are listed in parentheses for comparison.	
	The SHC values are in ($\hbar/e$) ($\Omega\cdot$cm)$^{-1}$.}
	\label{tb:tb1}
	\begin{tabular}{ccccc}
		\hline
		\hline
		Structure  &SHC($\Delta\mu$=0) & SHC$_{\max}$   &SHA($\Delta\mu$=0)  &SHA$_{\max}$ \\
		\hline
		$\alpha$-W  &$-785$    &$-1,206$   &$-0.02$ ($> -0.07$)\textsuperscript{\emph{a}}   &$-0.03$ \\
		$\beta$-W   &$-1,255$  &$-1,565$   &$-0.32$ ($-0.33$, $-0.40$)\textsuperscript{\emph{b}}  &$-0.47$ \\
		\hline
	\end{tabular}\\
	\begin{flushleft}
		\textsuperscript{\emph{a}} Reference \onlinecite{Pai2012}.\\
		\textsuperscript{\emph{b}} Reference \onlinecite{Pai2012, Hao2015b}.\\
	\end{flushleft}
\end{table}

\begin{figure} [tbp]
\centering
 \includegraphics[width=0.7\textwidth]{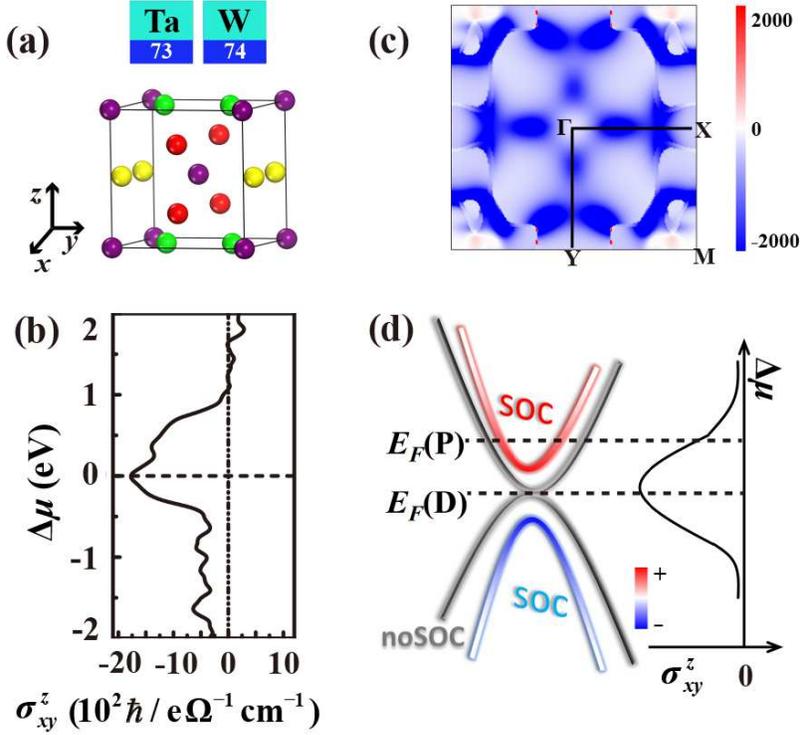}\\
 \caption{(Color online) (a) A15 crystal structure of $\beta$-W consisting of four nonequivalent sites at the center or corners (purple spheres) and at three atomic chains along the $x$ (yellow spheres), $y$- (green spheres), and $z$ (red spheres) directions, respectively.
 (b) Spin Hall conductivity, $\sigma_{xy}^z$, versus Fermi level position, $\Delta\mu$ for
 $\beta$-W alloyed with Tantalum substituting a host atom along the $x$-direction.
 (c) $\bf{k}_\|$-resolved spin Berry curvature in the 2D BZ ($k_x$, $k_y$) at $k_z$ = 0 and for Fermi level position at 0 eV.
 (d) Schematic diagram of the band structure around $\Gamma$ point without SOC (gray bands) and
 with SOC (red and blue band) where the red and blue color denotes bands with positive and negative spin berry curvature, respectively. $E_F(P)$ and $E_F(D)$ denote the Fermi level positions for pristine $\beta$-W and Ta-doped $\beta$-W,respectively.
 We also show schematically the variation of the SHC versus Fermi level position.}
 \label{fig:Figure3}
\end{figure}

\textit{Alloying $\beta$-W with Tantalum---}
In order to corroborate our prediction of the large enhancement of SHC via Ta alloying [see Fig. S3(b) in Supplementary Information\cite{support}], we have carried {\it ab initio} calculations of W$_{1-x}$Ta$_x$ based on $\beta$-W. Since the atomic number of Tantalum ($Z_{Ta}$ = 73) is smaller by one unit than that of W ($Z_{W}$ = 74), substitution of a single W atom by Ta yields one electron missing per unit cell. We considered a Ta atom on four nonequivalent sites in $\beta$-W [one for the corner/center site (purple spheres) and three sites on the $x$, $y$, and $z$ chains (red, yellow, and green spheres), respectively in Fig.~\ref{fig:Figure3}(a)]. Figure~\ref{fig:Figure3}(b) shows the SHC as a function of the Fermi level position,
$\Delta\mu$, for one Ta substitution on the chain along $x$-direction. Remarkably, the highest SHC value of $-1,773$ ($\hbar/e$) ($\Omega\cdot$cm)$^{-1}$ occurs at $\Delta\mu$ = 0 eV, which is a substantial enhancement of 41\% over the pristine $\beta$-W. Moreover, this result is consistent with the SHC at $\Delta\mu \approx -0.2$ eV in pristine $\beta$-W (Fig.~\ref{fig:Figure2}(a)). Note, that the SHC is robust with respect to small variations ($\approx \pm$ 0.2 eV) around $\Delta\mu = 0$ eV. Thus, the Ta alloying has a two-fold synergistic effect on enhancing the spin Hall angle via increasing the SHC and concurrently decreasing the longitudinal charge conductivity by disorder associated with alloying.

Figure~\ref{fig:Figure3}(c) shows the $\bf{k}_\|$-resolved spin Berry curvature in the 2D BZ ($k_x,k_y$) with $k_z = 0$ for the Fermi level position at $\Delta\mu = 0$ eV. One can see that the $\bf{k}_\|$-resolved Berry curvature is negative throughout the largest portion of the 2D BZ, which is very similar to that shown in Fig.~\ref{fig:Figure2}(c) for the pristine $\beta$-W at $\Delta\mu = -0.4$ eV. This strongly demonstrates that the effect of Ta alloying  is equivalent to rigidly lowering the Fermi level position in pristine $\beta$-W.

\textit{Mechanism of Ta-induced enhancement of SHC ---}
To elucidate the underlying mechanism of the Ta-induced enhancement of SHC, figure~\ref{fig:Figure3}(d) schematically illustrates the doubly-degenerate band around the $\Gamma$ without SOC (grey curves). The SOC splits this double-degeneracy with positive (in red) and negative (in blue) spin Berry curvature. If the Fermi level position ($\Delta\mu$) crosses either one of the SOC-split bands, the Berry curvature contributions of both sign tend to cancel each other. This is the case of pristine $\beta$-W where $E_{F}(P)$ crosses the upper band. On the other hand, if the Fermi level position lies in the spin-orbit coupling induced gap, the contribution from one sign of Berry curvature is excluded [positive in Fig.~\ref{fig:Figure3}(d)] resulting in an enhancement of the SHC. This is the case for Ta-doped $\beta$-W where $E_F(D)$ lies in the local gap along the high symmetry $\Gamma$-X direction [see Fig. S3(a) in Supplementary Information\cite{support}].

Electronic structure calculations are carried out for Ta-alloyed $\beta$-W where Ta substitutes one of the W atoms along the $y$- and $z$ chains and at center sites. We find that the behavior of SHC with Fermi level position is similar to that of Ta occupying a site along the $x$-chain shown in Fig.~\ref{fig:Figure3}(c) [see Fig. S3(b) in Supplementary Information\cite{support}], demonstrating that the intrinsic SHC of W$_{1-x}$Ta$_x$ is remarkably insensitive to the variations of local atomic environment.  It should be emphasized that the rigid-band model which shift the Fermi level inside the spin-orbit induced local bandgap can be used in various systems. Substitution of Ta for W will also cause large enhancement of spin Hall effect. Therefore, we provide new guidelines to tailor the intrinsic spin Hall effect by rigid-band model utilizing  alloying either with deficit or excessive valence electrons.

In summary, using first-principles calculations, we have revealed the physics origin of intrinsic SHC in both Tungsten phases. While $\beta$-phase exhibits about 60\% larger intrinsic spin Hall conductivity of than $\alpha$-W, much smaller longitudinal conductivity gives giant spin Hall angle. Moreover, we predict even more giangatic SHC in $\beta$-W via Ta alloying W$_{1-x}$Ta$_x$ ($x$ $\sim$ 12.5\%), where shift of Fermi level synergistically enhances SHC and reduces the longitudinal conductivity. Finally, we give a general mechanism to improve the intrinsic SHC: tuning the Fermi level of system to spin-orbit-coupling induced gap can induce large enhancement of spin Berry curvature integral. Our results pave an intriguing way towards enhancing the magnitude of SHA and SOT in ferromagnetic/heavy metal bilayers using Fermi level variations. Our predictions may motivate further experimental studies of the crucial effect of Ta-alloying in $\beta$-W so as to enhance the SHA and hence optimize the spin-orbit torque efficiency.

We acknowledge the support of the NSF Grant No. ERC TANMS-1160504 and Ministry of Science and Technology of China (Grant No. 2016YFA0301001), and the National Natural Science Foundation of China (Grants No. 11674188 and 11334006). This work is supported in part by the Beijing Advanced Innovation Center for Future Chip (ICFC). Work at Ulsan is supported by Creative Materials Discovery Program through the National Research Foundation of Korea (2015M3D1A1070465).

\end{document}